\documentclass[10pt, conference]{IEEEtran}

\usepackage{cite}
\usepackage{amsmath,amssymb,amsfonts}
\usepackage{bbm,epsfig,bm}
\usepackage{algorithmic}
\usepackage{graphicx}
\usepackage{textcomp}
\usepackage{xspace}
\usepackage[normalem]{ulem}
\usepackage[binary-units=true]{siunitx}
\usepackage[usenames]{xcolor}
\usepackage{booktabs}
\usepackage{arydshln} 
\def\BibTeX{{\rm B\kern-.05em{\sc i\kern-.025em b}\kern-.08em
    T\kern-.1667em\lower.7ex\hbox{E}\kern-.125emX}}
\usepackage[capitalise]{cleveref}
\usepackage{tikz}
\usetikzlibrary{calc,shapes,arrows,positioning}

\usepackage{subcaption}
\usepackage{tcolorbox}
\usepackage{balance}

\usepackage{listings}
\definecolor{dkgreen}{rgb}{0,0.5,0}
\definecolor{lessdkgreen}{rgb}{0,0.6,0}
\definecolor{dkred}{rgb}{0.5,0,0}
\definecolor{gray}{rgb}{0.5,0.5,0.5}

\definecolor{figgray}{rgb}{0.96,0.96,0.96}
\definecolor{figgreen}{rgb}{0.81,0.92,0.56}
\definecolor{figblue}{rgb}{0.80,0.90,0.99}
\definecolor{figdkblue}{rgb}{0.42,0.55,0.72}
\definecolor{figred}{rgb}{0.97,0.81,0.81}
\definecolor{figdkred}{rgb}{0.71,0.33,0.32}

\lstdefinestyle{cstyle}{
language=c,
basicstyle=\ttfamily\bfseries\scriptsize,
  morekeywords={virtualinvoke},
  keywordstyle=\color{blue},
  ndkeywordstyle=\color{red},
  commentstyle=\color{dkred},
  stringstyle=\color{dkgreen},
  numbers=left,
  breaklines=true,
  numberstyle=\ttfamily\footnotesize\color{gray},
  stepnumber=1,
  numbersep=10pt,
  backgroundcolor=\color{white},
  tabsize=4,
  showspaces=false,
  showstringspaces=false,
  xleftmargin=.23in,
  captionpos=b,
  escapeinside={$}{$},
  print
}
\lstdefinestyle{cinlinestyle}{
language=c,
basicstyle=\ttfamily\bfseries\footnotesize,
  morekeywords={virtualinvoke},
  keywordstyle=\color{blue},
  ndkeywordstyle=\color{red},
  commentstyle=\color{dkred},
  stringstyle=\color{dkgreen},
  escapeinside={$}{$},
  print
}

\lstdefinelanguage
    [hexrays]{Assembler}
    [x86masm]{Assembler}
    {morekeywords={retn,rbp,rsp},deletendkeywords={dword,ptr,short},morendkeywords={var1,var2}
    }

\lstdefinestyle{hexrays}{
  language=[hexrays]{Assembler},
  basicstyle=\ttfamily\bfseries\scriptsize,
  keywordstyle=\color{blue},
  ndkeywordstyle=\color{lessdkgreen},
  commentstyle=\color{dkred},
  print
}

\newcommand{\eg}[0]{e.g., }
\newcommand{\ie}[0]{i.e., }
\newcommand{\etal}[0]{et al.}

\newcommand{\GH}{{\sc GitHub}\xspace}
\newcommand{\debin}{{\sc Debin}\xspace}

\newcommand{\approach}[0]{\textsc{DIRE}\xspace}
\newcommand{\expandedname}[0]{\textbf{D}ecompiled \textbf{I}dentifier
  \textbf{R}enaming \textbf{E}ngine\xspace}

\newcommand{\successpct}[0]{74.3\%\xspace}
\newcommand{\numbinaries}[0]{164,632\xspace}
\newcommand{\numfunctions}[0]{3,195,962\xspace}

\newcommand\tk{\ensuremath{x}}

\newcommand\lt[1]{{\lstinline[style=cinlinestyle]!#1!}}
\renewcommand{\tt}{\texttt}

\begin{document}

\title{\approach: A Neural Approach to Decompiled Identifier Naming}

\author{
  \IEEEauthorblockN{
  	Jeremy Lacomis\IEEEauthorrefmark{1},
  	Pengcheng Yin\IEEEauthorrefmark{1},
	Edward J. Schwartz\IEEEauthorrefmark{2},
	Miltiadis Allamanis\IEEEauthorrefmark{3}, \\
	Claire Le~Goues\IEEEauthorrefmark{1},
	Graham Neubig\IEEEauthorrefmark{1},
	Bogdan Vasilescu\IEEEauthorrefmark{1}
  }
  \vspace{0.2cm}
  \IEEEauthorblockA{
    \IEEEauthorrefmark{1}Carnegie Mellon University.
    	\{jlacomis, pcyin, clegoues, gneubig\}@cs.cmu.edu; vasilescu@cmu.edu
  }
  \IEEEauthorblockA{
    \IEEEauthorrefmark{2}Carnegie Mellon University Software Engineering Institute.
    	eschwartz@cert.org
  }
  \IEEEauthorblockA{
    \IEEEauthorrefmark{3}Microsoft Research.
    	miallama@microsoft.com
  }
}
\maketitle

\thispagestyle{plain}
\pagestyle{plain}

\begin{abstract}
  The \emph{decompiler} is one of the most common tools for examining binaries without corresponding source code.
  It transforms binaries into high-level code, reversing the compilation process.
  Decompilers can reconstruct much of the information that is lost during the compilation process (\eg structure and type information).
  Unfortunately, they do not reconstruct semantically meaningful variable names, which are known to increase code understandability.
  We propose the \expandedname (\approach), a novel probabilistic technique for variable name recovery that uses both lexical and structural information recovered by the decompiler.
  We also present a technique for generating corpora suitable for training and evaluating models of decompiled code renaming, which we use to create a corpus of \numbinaries unique x86-64 binaries generated from C projects mined from \textsc{GitHub}.\footnote{\label{datanote}Data available at https://doi.org/10.5281/zenodo.3403077}
  Our results show that on this corpus \approach can predict variable names identical to the names in the original source code up to \successpct of the time.
\end{abstract}

\section{Introduction}
\emph{Software reverse engineering} is the problem of understanding the behavior of a program without having access to its source code.
Reverse engineering is often used to predict the behavior of malware~\cite{Yakdan2015,Yakdan2016,Durfina2013}, discover vulnerabilities~\cite{Yakdan2015,VanEmmerik2007,Schwartz2013}, and patch bugs in legacy software~\cite{VanEmmerik2007,Schwartz2013}.
For malware and malicious botnets, reverse engineering enables understanding and response, and helps identify and patch infection vectors.
For example, by reverse engineering the Torbig botnet (which caused 180K infections and collected \SI{70}{\giga\byte} of credit card/bank account information), responders were able to predict future domain names that bots would contact, and redirect the bots to servers under the responders' control~\cite{StoneGross2009}.
Reverse engineering can also help identify who created a piece of malware, as was done for the Uroburos rootkit~\cite{Uroburos2014} (which captured files and network traffic while propagating over networks of companies and public authorities), and estimate the extent of infection~\cite{Rossow2013}.

One of the main tools reverse engineers use to inspect programs is the \emph{disassembler}---a tool that translates a binary to low-level assembly code.
Disassemblers range from simple tools like GNU Binutils' objdump~\cite{objdump}, to more advanced tools like IDA~\cite{ida}, which can be used interactively and have more sophisticated features.
However, reasoning at the assembly level requires considerable cognitive effort even with these advanced features~\cite{VanEmmerik2007,Schwartz2013,Yakdan2016}.
More recently, reverse engineers are employing \emph{decompilers} such as Hex-Rays~\cite{hexrays} and Ghidra~\cite{ghidra}, which reverse compilation by translating the output of disassemblers into code that resembles high-level languages such as C, to reduce the cognitive burden of understanding assembly code.
These state-of-the-art tools are able to use program analysis and heuristics to reconstruct information about a program's variables, types, functions, and control flow structure.

Even though decompiler output is more understandable than raw assembly, decompilation is often incomplete.
Compilers discard source-level information and lower its level of abstraction in the interest of binary size, execution time, and even obfuscation.
Comments, variable names, user-defined types, and idiomatic structure are all lost at compile time, and are typically unavailable in decompiler output.
In particular, variable names, which are highly important for code comprehension and readability~\cite{Gellenbeck1991,Lawrie2006}, become nothing more than arbitrary placeholders such as \lt{VAR1} and \lt{VAR2}.

In this work, we present \approach (\expandedname), a novel neural network approach for assigning meaningful names to variables in decompiled code (\cref{sec:model}).
To build \approach, we rely on two key insights.
Our first insight is that software is \emph{natural}, \ie programmers tend to write similar code and use the same variable names in similar contexts~\cite{Hindle2012,Devanbu2015}.
Therefore, because of this repetitiveness, if given a large enough training corpus one can \emph{learn} appropriate variable names for a particular context.

Prior approaches exist to predict natural variable names from both source code~\cite{JSNice,JSNaughty,Context2Name,Alon2018} and compiled executables~\cite{Jaffe2018,DEBIN}.
However, approaches to predict variable names from executables either operate directly on the binary semantics~\cite{David2019,DEBIN}, or on the lexical output of the decompiler~\cite{Jaffe2018}.
The former ignores the rich abstractions that modern decompilers are able to recover.
The latter is an improvement, but a lexical program representation is by its very nature sequential, and lacks rich structural information that could be used to improve predictions.
In contrast, \approach uses the extended context provided by the decompiler's internal abstract syntax tree (AST) representation of the decompiled binary, which encodes additional \emph{structural} information.

To train such models, one needs training data that specifies what names are natural in what contexts.
Our second key insight is that unlike other domains, where creating training data often requires manual curation (\eg machine translation~\cite{koehn2009statistical}), it is possible to \emph{automatically} generate large amounts of training data for identifier name prediction, To that end, we mine open-source C code from \GH, compile it \emph{with debugging information} such that the binaries preserve the original names, and decompile those binaries so that the output contains the original names.
We then strip the debug symbols, decompile the binary again, and identify the alignment between the identifiers in the two versions of the decompiler outputs.
While this is conceptually straight-forward, the two outputs are not simply $\alpha$-renamings, making the process of calculating these alignments far from trivial.
Prior work identified alignments based entirely on heuristics~\cite{Jaffe2018}.
In contrast, we observe that the set of instruction addresses that access each variable uniquely identifies that variable, and this can be used to generate accurate alignments (\cref{sec:corpus}).

With these insights we train and evaluate \approach on a large dataset of C code mined from \GH, showing that we can predict variable names identical to those chosen by the original developer up to \successpct of the time.
In short, we contribute:
\begin{itemize}
\item \expandedname (\approach), a technique for assigning meaningful names to decompiled variables that outperforms previous approaches.
\item A novel technique for generating corpora suitable for training both lexical and graph-based probabilistic models of variable names in decompiled code.
\item A dataset of \numfunctions decompiled x86-64 functions and parse trees annotated with gold-standard variable names.\footnotemark[1]
\end{itemize}

\section{Background}
\label{sec:background}

Before diving into the technical details of our approach, we start with some background on decompilation, statistical models of source code, and the two particular classes of deep learning models we rely on, recurrent neural networks (RNNs) and gated-graph neural networks (GGNNs).

\subsection{Decompilation}
\label{sec:bg:decompilation}

At a high level, a compiler generates binaries from source using a pipeline of processing stages, and decompilers try to reverse this pipeline using various techniques~\cite{Katz2018}.
Typically, a binary is first passed through a platform-specific \emph{disassembler}.
Next, assembly code is typically \emph{lifted} to a platform-independent intermediate representation (IR) using a binary-to-IR lifter.
The next stage is the heart of the decompiler, and is where a number of program analyses are used to recover variables, types, functions and control flow abstractions, which are ultimately combined to reconstruct an abstract syntax tree (AST) corresponding to an idiomatic program.
Finally, a code generator converts the AST to the decompiled output.

\begin{figure}
  \tikzstyle{codebox} = [draw, rectangle, rounded corners, inner sep=2pt,
  inner ysep=0pt]
  \begin{tikzpicture}[every edge/.append style = { ->, thick, >=stealth,
  line width= 0.5pt}, baseline={(whileloop.base)}]
    \node [codebox] (forloop) {%
      \begin{minipage}{0.4\columnwidth}
        \begin{lstlisting}[style=cstyle, showlines=true]
int i;
for (i=0; i<10; i++)
{
  z+=i;
}
        \end{lstlisting}
      \end{minipage}
    };
    \node [codebox, below = .2cm of forloop] (whileloop) {%
      \begin{minipage}[t]{0.4\columnwidth}
        \begin{lstlisting}[style=cstyle]
int n=0;
while (n<10) {
  x+=n;
  n++;
}
        \end{lstlisting}
      \end{minipage}
    };
    \node [codebox, anchor=center ] (assembly) at (4.6, -1) {%
      \begin{minipage}{0.53\columnwidth}
        \begin{lstlisting}[style=hexrays, escapechar=!]
487: var1 = dword ptr -8
487: var2 = dword ptr -4
     ;...
492:       mov  [rbp+var2], 0
499:       jmp  loc_4A5
49B: loc_49B:
49B:       mov  eax, [rbp+var2]
49E:       add  [rbp+var1], eax
4A1:       add  [rbp+var2], 1
4A5: loc_4A5:
4A5:       cmp  [rbp+var2], 9
4A9:       jle  loc_49B
        \end{lstlisting}
      \end{minipage}
    };
    \draw (whileloop.east) edge (assembly.west|-whileloop.east);
    \draw (forloop.east) edge (assembly.west|-forloop.east);
  \end{tikzpicture}
  \caption{Two different C loops that compile to the same assembly code.
    Note the normalized structure and names.\vspace{-.2cm}
    \label{fig:loops}}
\end{figure}
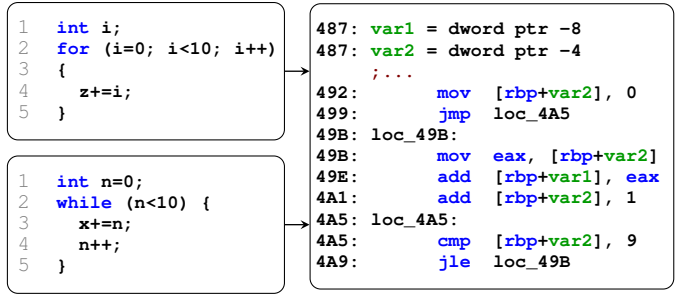

\looseness=-1
Decompilation is more difficult than compilation, because each stage of a compiler loses information about the original program.
For example, the lexing/parsing stage of the compiler does not propagate code comments to the AST.
Similarly, converting from the AST to IR can lose additional information.
This loss of information allows multiple distinct source code programs to compile to the same assembly code.
For example, the two loops in \cref{fig:loops} are reduced to the same assembly instructions.
The decompiler cannot know which source code was the original, but it does try to generate code that is \emph{idiomatic}, using heuristics to increase code readability.
For example, high-level control flow structures such as \lt{while} loops are preferred over \lt{goto} statements.

The choice of which code to generate is largely heuristic, but can be informed by the inclusion of DWARF debugging information~\cite{DWARF}.
This debugging information, which can optionally be generated at compile-time, greatly assists the decompiler by identifying function offsets, types of variables, identifier names, and user-defined structures and unions.

\subsection{Statistical Models of Source Code}

A wide variety of statistical models for representing source code have been proposed based on the \emph{naturalness} of software~\cite{Hindle2012,Devanbu2015}.
This key property states that source code is highly repetitive given context, and is therefore predictable.
Statistical models capture the implicit knowledge hidden within code, and apply it to build new software development tools and program analyses, \eg for code completion, documentation generation, and automated type annotation~\cite{allamanis2018survey}.

Predicting variable names is no exception.
Work has shown that statistical models trained on source code corpora can predict descriptive names for variables in a previously-unseen program, given the contextual features of the code the variable is used in.
These naming models can help to distill coding conventions~\cite{Allamanis2014} or analyze obfuscated code~\cite{JSNice,JSNaughty}.
Several classes of statistical models have been used for renaming, including $n$-grams~\cite{Allamanis2014, JSNaughty}, conditional random fields (CRFs)~\cite{JSNice}, and deep learning models~\cite{Allamanis2015suggesting, Allamanis2018, alon2019code2seq}.

Two recent approaches aim to suggest informative variable names in decompiled code.
Our prior work~\cite{Jaffe2018} proposed a lexical $n$-gram-based machine translation model that operates on decompiler output.
That approach used heuristics to align variables in the decompiler output and original source, which are needed for training, and is able to exactly recover 12.7\% of the original names in the test set.
Contemporaneously, He \etal~\cite{DEBIN} proposed a two-step approach that operates on a stripped binary rather than the decompiler output.
First, the authors predict whether a low-level register or a memory offset maps to a variable at the source-level.
Then, using structured prediction with CRFs, they predict names and types for the mapped variables.
63.5\% of the variables in the test set for which the first step succeeded could be recovered exactly.

\subsection{Neural Network Models}
\label{sec:nnets}

Our approach builds on two advances in statistical models for representing source code: recurrent neural networks (RNNs) and gated-graph neural networks (GGNNs).

\subsubsection{Recurrent Neural Networks}
\label{sec:rnn}

RNNs are networks where connections between nodes form a sequence~\cite{rumelhart1988learning}.
They are typically used to process sequences of inputs by reading in one element at a time, making them well-suited to sequences, such as source code tokens.
In this work, we use long short-term memory (LSTM) models~\cite{hochreiter1997long}, a variant of RNNs widely used in text processing.

Formally, an LSTM has the following structure: given a sequence of tokens $\{ \tk_i \}_{i=1}^{n}$, an LSTM $\overrightarrow{f}_{\textrm{LSTM}}$ processes them in order, maintaining a hidden state $\bm{\overrightarrow{h_i}}$ for each subsequence up to token $\tk_i$ using the recurrent function $\bm{\overrightarrow{h_i}} = \overrightarrow{f}_\textrm{LSTM}(\textrm{emb}(\tk_i), \bm{\overrightarrow{h_i}_{-1}})$, where $\textrm{emb}(\cdot)$ is an embedding function mapping $\tk_i$ into a learnable vector of real numbers.

As we will elaborate later in~\cref{sec:model}, we use two types of LSTMs in~\approach: encoding LSTMs and decoding LSTMs.
An encoder LSTM reads the input sequence (\eg a sequence of source code tokens, as in~\cref{sec:neural_model:encoder:lexical}) and encodes it into continuous vectors; while a decoder LSTM takes these vectors and generates the output sequence (\eg the sequence of predicted names for all identifiers, as in~\cref{sec:neural_model:decoder}).

\subsubsection{Gated-Graph Neural Networks}
\label{sec:ggnn}

While LSTMs are useful for modeling sequences, they do not capture any additional structural information.
Within the decompilation task, structured information provided by the AST is a natural information source about choice of variable names.
For this purpose, we also employ structural encoding of the code using GGNNs, a class of neural models that map \emph{graphs} to outputs~\cite{scarselli2009graph,li2015gated}.
At a high level, GGNNs are neural networks over directed graphs.
Initially, we associate each vertex with a learned or computed hidden state containing information about the vertex.
GGNNs compute representations for each node based on the initial node information and the graph structure.

Formally, let $\mathcal{G}=\langle V, E \rangle$ be a directed graph describing our problem, where $V=\{v_i\}$ is the set of vertices and $E= \{ (v_i \mapsto v_j, \mathcal{T}) \}$ is the set of typed edges.
Let $\mathcal{N}_{\mathcal{T}}(v_i)$ denote the set of vertices adjacent to $v_i$ with edge type $\mathcal{T}$.
In a GGNN, each vertex $v_i$ is associated with a state $\bm{h}^{g}_{i, t}$ indexed by a time step $t$.
At each time step $t$, the GGNN updates the state of all nodes in $V$ via neural message passing (NMP)~\cite{gilmer2017neural}.
Concurrently for each node $v_i$ at time $t$, NMP is performed as follows: First, for each $v_{j} \in \mathcal{N}_{\mathcal{T}}(v_i)$ we compute a message vector $\bm{m}_{\mathcal{T}}^{v_j \mapsto v_i} = \bm{W}_{\mathcal{T}} \cdot \bm{h}^{g}_{j, t - 1}$, where $\bm{W}_{\mathcal{T}}$ is a type-specific weight matrix.
Then, all $\bm{m}_{\mathcal{*}}^{v_* \mapsto v_i}$ are aggregated, and summarized into a single vector $\bm{x}_i^{g}$ via element-wise mean (pooling):
\[
  \bm{x}^{g}_i = \mathsf{MeanPool}(\{ \bm{m}^{v_j \mapsto v_i}_{\mathcal{T}} : v_j \in \mathcal{N}_{\mathcal{T}}(v_i), \forall \mathcal{T} \}).
\]
Finally, the state of every node $v_i$ is updated using a nonlinear activation function $f$: $ \bm{h}^{g}_{i, t} = f(\bm{x}^{g}_i, \bm{h}^{g}_{i, t - 1})$.
GGNNs use a Gated Recurrent Unit (GRU) update function, $f_\textrm{GRU}(\cdot)$, introduced by Cho \etal~\cite{Cho2014LearningPR}.
By repeatedly applying NMP for $T$ steps, each node's state gradually represents information about that node and its \emph{context} within the graph.
The computed states can then be used by a decoder, similarly to the LSTM-based decoder architectures.
As in LSTMs, all GGNN parameters (parameters of $f_\textrm{GRU}(\cdot)$ and the $\bm{W}_{\mathcal{T}}$s) are optimized along with the rest of the model.

\section{The DIRE Architecture}
\label{sec:model}

\newcommand{\gear}[6]{%
  (0:#2)
  \foreach \i [evaluate=\i as \n using {\i-1)*360/#1}] in {1,...,#1}{%
    arc (\n:\n+#4:#2) {[rounded corners=.5pt] -- (\n+#4+#5:#3)
    arc (\n+#4+#5:\n+360/#1-#5:#3)} --  (\n+360/#1:#2)
  }%
  (0,0) circle[radius=#6]
}
  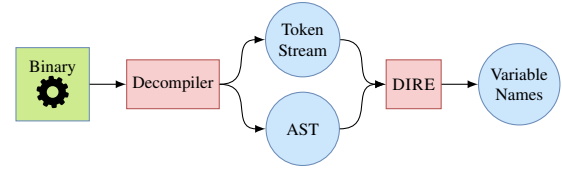
\begin{figure}
  \centering
  \tikzstyle{data} = [circle, draw=figdkblue, fill=figblue, minimum size=1.5cm, align=center]
  \tikzstyle{stage} = [rectangle, draw=figdkred, fill=figred, minimum size=1cm]
  \begin{tikzpicture}[node distance = .75cm, auto, scale=0.65, every
      node/.style={transform shape}]
    \node [data] (tok) {Token\\Stream};
    \node [data, below = .3cm of tok] (ast) {AST};
    \node [stage, left = 1.7cm of tok] (decomp) at ($(tok)!0.5!(ast)$) {Decompiler};
    \node [stage, right = 1.7cm of tok] (dire) at ($(tok)!0.5!(ast)$) {\approach};
    \node [data, rectangle, fill=figgreen, left = of decomp] (bin) {Binary\\
      \begin{tikzpicture}[scale=0.3]
        \fill[even odd rule] \gear{10}{.9}{1.1}{12}{3.4}{.4};
      \end{tikzpicture}
    };
    \node [data, right = of dire] (output) {Variable\\Names};
    \draw[->,>=latex] (bin.east) -- (decomp.west);
    \draw[->,>=latex] (decomp.east) .. controls ([xshift=.75cm] decomp.east) and
    ([xshift=-.75cm] tok.west) .. (tok.west);
    \draw[->,>=latex] (decomp.east) .. controls ([xshift=.75cm] decomp.east) and
    ([xshift=-.75cm] ast.west) .. (ast.west);
    \draw[->,>=latex] (tok.east) .. controls ([xshift=.75cm] tok.east) and
    ([xshift=-.75cm] dire.west) .. (dire.west);
    \draw[->,>=latex] (ast.east) .. controls ([xshift=.75cm] ast.east) and
    ([xshift=-.75cm] dire.west) .. (dire.west);
    \draw[->,>=latex] (dire.east) -- (output.west);
  \end{tikzpicture}
  \caption{High-level overview of our approach.\label{fig:arch}
  }
\end{figure}

We start with an overview of our approach, then dive into the technical details of each component.

\subsection{Overview}

We designed \approach to work on top of a decompiler as a plugin that can automatically suggest more informative variable names.
We use Hex-Rays, a state-of-the-art industry decompiler, though our approach is not fundamentally coupled to Hex-Rays and can be adapted to other decompilers.

\cref{fig:arch} gives a high-level overview of our workflow.
First, a binary is passed to a decompiler, which decompiles each function in the binary.
For each function, our plugin traverses the AST, inserting placeholders at variable nodes.
This produces two outputs: the AST and the tokenized code.
These outputs are provided as input to our neural network model, \approach, which generates unique variable names for each placeholder in the input.
The decompiler output can then be rewritten to include the suggested variable names.

\begin{figure*}[tb]
  \centering
  \includegraphics[width=\textwidth]{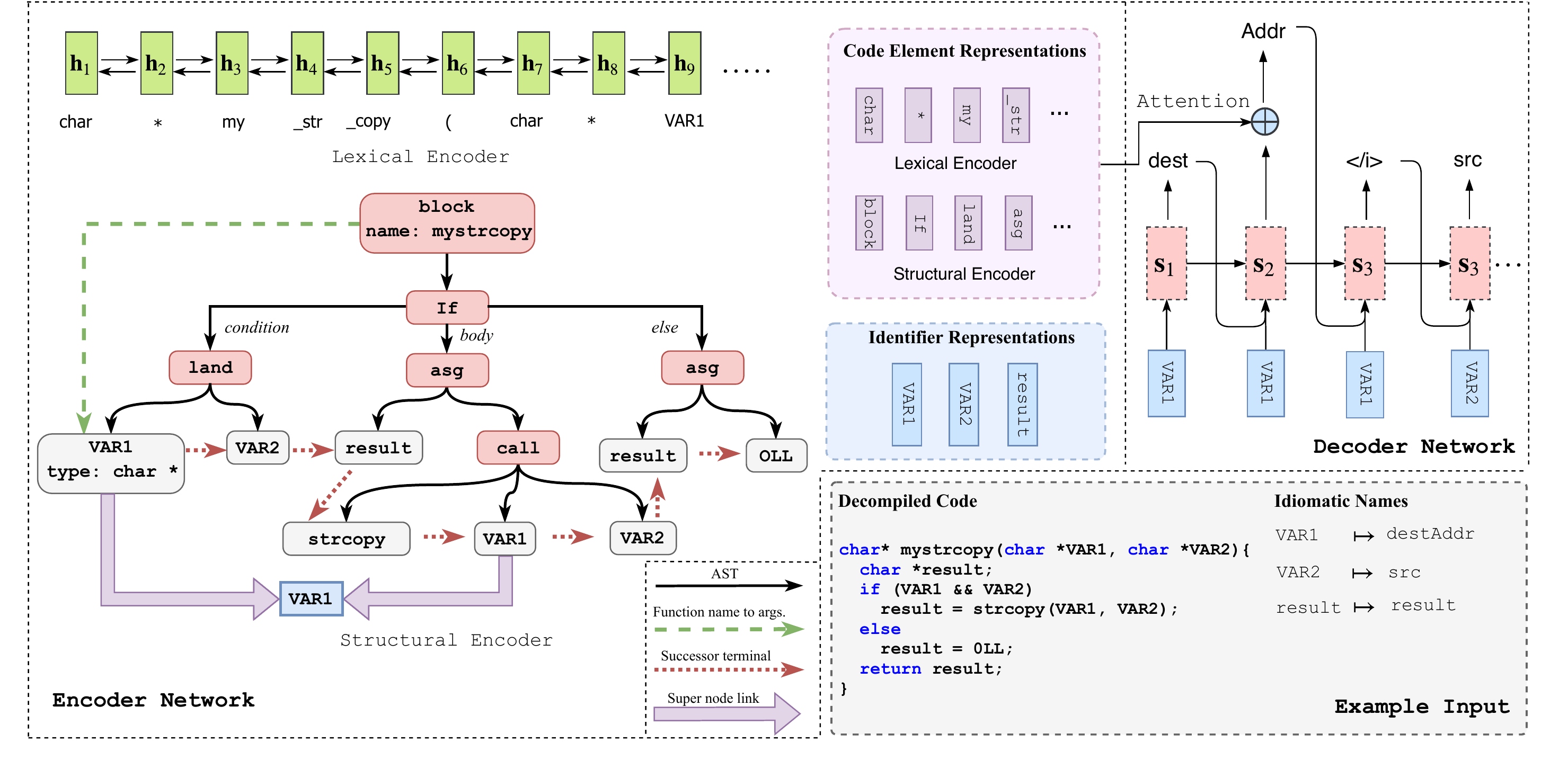}
  \caption{Overview of \approach's neural architecture.
    For clarity, we omit the data-flow links in the AST in the structural encoder.}
  \label{fig:neural_model}
\end{figure*}

\cref{fig:neural_model} gives an overview of the neural architecture.
\approach follows an encoder-decoder architecture: An \emph{encoder} neural network (\cref{sec:neural_model:encoder}) first encodes the decompiler's output---both the sequence of decompiled code tokens and its internal AST---and computes distributed representations (\ie real-valued vectors, or \emph{embeddings}) for each identifier and code element.
These encoded representations are then consumed by a \emph{decoder} neural network (\cref{sec:neural_model:decoder}) that predicts meaningful names for each identifier based on the contexts in which it is used.

The key takeaway is that \approach uses both lexical information obtained from the tokenized code as well as structural information obtained from the corresponding ASTs.
This is achieved by using two encoders---a \emph{lexical encoder} (\cref{sec:neural_model:encoder:lexical}) and a \emph{structural encoder} (\cref{sec:neural_model:encoder:structural})---to separately capture the lexical and structural signals in the decompiled code.
As we will show, this combination of lexical and structural information allows \approach to outperform techniques that rely on lexical information alone~\cite{Jaffe2018}.

\subsection{The Encoder Network}
\label{sec:neural_model:encoder}

Each encoder network in \approach outputs two sets of representations:
\begin{itemize}
\item A \emph{code element representation} for each element in the decompiler's output.
  Depending on the type of the encoder, a code element will either be a token in the surface code (for the lexical encoder), or a node in the decompiler's internal AST (for the structural encoder).
  \item An \emph{identifier representation} for each unique identifier defined in the input binary, which is a real-valued vector that represents the identifier in the neural network.
\end{itemize}
The lexical and structural representations are then merged to generate a unified encoding of the input binary (dashed boxes in~\cref{fig:neural_model}).
By computing separate representations for code elements and identifiers, the \approach decoder can better incorporate the contextual information in the encodings of individual code elements to improve name predictions for the different identifiers; see~\cref{sec:neural_model:decoder}.

\subsubsection{Lexical Code Encoder}
\label{sec:neural_model:encoder:lexical}

The lexical encoder sequentially encodes the tokenized decompiled code, projecting each token $\tk_i$ into a fixed-length vector encoding $\bm{\tk}_i$.
Specifically, the lexical encoder uses the sub-tokenized code as the input, where a complex code token (\eg the function name \lt{mystrcopy}) is automatically broken down into sub-pieces (\eg \lt{my}, \lt{str}, and \lt{copy}) using SentencePiece~\cite{kudo2018sentencepiece}, based on sub-token frequency statistics.
Sub-tokenization reduces the size of the encoder's vocabulary (and thus its training time), while also mitigating the problem of rare or unknown tokens by decomposing them into more common subtokens.
We treat the placeholder and reserved variable names (\eg \lt{VAR1}, \lt{VAR2}, and the decompiler-inferred name \lt{result}) in the decompiler's output as special tokens that should not be sub-tokenized.

\approach implements the lexical encoder using LSTMs (described in \cref{sec:rnn}).
We use a bidirectional LSTM: The forward network $\overrightarrow{f}_{\textrm{LSTM}}$ processes the tokenized code $\{ \tk_i \}_{i=1}^{n}$ sequentially.
The backward LSTM processes the input tokenized code in backward order, producing a backward hidden state $\overleftarrow{\bm{h}_i}$ for each token $\tk_i$.
Intuitively, a bidirectional LSTM captures informative context around a particular variable both before and after its sequential location.

\emph{Element Representations} We encode a token $\tk_i$ by concatenating its asssociated state vectors, \ie $\bm{\tk}_i = [ \overrightarrow{\bm{h}_i} : \overleftarrow{\bm{h}_i}]$, a common strategy in source code representations using LSTMs~\cite{allamanis2018survey}.
For a particular token $\tk_i$ we compute the forward (resp. backward) representation using both its embedding and the hidden states of its preceding (resp. succeeding) tokens.
This is important because the resulting encoding $\bm{\tk}_i$ captures both the local and contextual information of the current token and its surrounding code.

To compute the \emph{identifier} representation $\bm{v}$ for each unique identifier $v$, we collect the set of subtoken representations $\mathcal{H}_{v}$ of $v$, and perform an element-wise mean over $\mathcal{H}_{v}$ to get a fixed-length representation: $ \bm{v} = \mathsf{MeanPool}(\mathcal{H}_{v})$.

\subsubsection{Structural Code Encoder}
\label{sec:neural_model:encoder:structural}

The lexical encoder only captures sequential information in code tokens.
To also learn from the rich structural information available in the decompiler AST, \approach employs a gated-graph neural network (GGNN) structural encoder over the AST (\cref{sec:ggnn}).
This requires a mechanism to compute initial node states, as well as design choices of which AST edges to consider in the node encoding:

\paragraph{Initial Node States} The initial state of a node $v_i$, $\bm{h}^{g}_{i, t=0}$ is computed from three separate embedding vectors, each capturing different types of information of $v_i$:
1) An embedding of the node's syntactic type (\eg the root note in the AST in \cref{fig:neural_model} has the syntactic type \lt{block}).
2) For a node that represents data (\eg variables, constants) or an operation on data (\eg mathematical operators, type casts, function calls), an embedding of its data type, computed by averaging the embeddings of its subtokenized type.
For instance, the variable node \lt{VAR1} in \cref{fig:neural_model} has the data type \lt{char *}; its embedding is computed by averaging the embeddings of the type subtokens \lt{char} and \lt{*}.
3) For named nodes, an embedding of the node's name (\eg the root node in \cref{fig:neural_model} has a name \lt{mystrcopy}), computed by averaging the embeddings of its content subtokens.
The initial state $\bm{h}^{g}_{v, t=0}$ is then derived from a linear projection of the concatenation of the three separate embedding vectors. For nodes without a data type or name, we use a zero-valued vector as the respective embedding.

\paragraph{Graph Edges}
Our structural encoder uses different types of edges to capture different types of information in the AST.
Besides the simple {\it parent-child} edges (solid arrows in the AST in \cref{fig:neural_model}) in the original AST, we also augment it with additional edges~\cite{Allamanis2018}:
\begin{itemize}
\item We add an {\color[rgb]{0.467,0.667,0.357} edge} from the root \lt{block} node containing the function name to each identifier node.
  The function name can inform names of identifiers in its body.
  In our running example the two arguments \lt{VAR1} and \lt{VAR2} defined in the \lt{mystrcopy} function might indicate the source and destination of the copy.
  This type of link (``Function name to args.'' in \cref{fig:neural_model}) captures these naming dependencies.
  \item To capture the dependency between neighboring code, we add an {\color[rgb]{0.686,0.290,0.278} edge} from each terminal node to its lexical successor (``Successor terminal'').
  \item To propagate information among all mentions of an identifier, we add a virtual ``supernode'' (rectangular node labeled \lt{VAR1}) for each unique identifier $v_i$, and {\color[rgb]{0.545,0.408,0.612} edges} from mentions of $v_i$ to the supernode (``Super node link'')~\cite{gilmer2017neural}.
  \item Finally, we add a reverse edge for all edge types defined above, modeling bidirectional information flow.
\end{itemize}

\paragraph{Representations} For the \emph{element} representation, we use the final state of the GGNN for node $n_i$, $\bm{h}^{g}_{i, T}$, as its representation: $\bm{n}_i = \bm{h}^{g}_{i, T}$ (the recurrent process unrolls $T$ times; $T=8$ for all our experiments).
For the \emph{identifier} representation for each unique identifier $v_i$, its representation $\bm{v}_i$ is defined as the final state of its supernode as the encoding of $v_i$.
Since the supernode has bidirectional connections to all the mentions of $v_i$, its state is computed using the states of all its mentions.
Therefore, $\bm{v}_i$ captures information about the usage of $v_i$ in different occurrences.

\subsubsection{Combining Outputs of Lexical and Structural Encoders}
The lexical and the structural encoders output a set of representations for each identifier and code element.
In the final phase of encoding, we combine the two sets of outputs.
Code elements are combined by unioning the lexical set (of code tokens) and structural set (of AST nodes) of element representations as the final encoding of each input code element; identifiers are combined by merging the lexical and structural representations of each identifier $v$ using a linear transformation as its representation.

\subsection{The Decoder Network}
\label{sec:neural_model:decoder}

The decoder network predicts names for identifiers using the representations given by the encoder.
As shown in \cref{fig:neural_model}, the decoder predicts names based on both the representations of identifiers, and contextual information in the encodings of code elements.
Specifically, as with the encoder, we assume an identifier name is composed of a sequence of sub-tokens (\eg \lt{destAddr} $\mapsto$ \lt{dest}, \lt{Addr}; see \cref{sec:neural_model:encoder:lexical}).

The decoder factorizes the task of predicting idiomatic names to a sequence of time-indexed decisions, where at each time step, it predicts a sub-token in the idiomatic name of an identifier.
For instance, the idiomatic name for \lt{VAR1}, \lt{destAddr}, is predicted in three time steps ($s_1$ through $s_3$) using sub-tokens \lt{dest}, \lt{Addr}, and \lt{</i>}, (the special token \lt{</i>} denoting the end of the token prediction process).
Once a full identifier name is generated, the decoder continues to predict other names following a pre-order traversal of the AST.
As we will elaborate in~\cref{sec:corpus}, not all identifiers in the decompiled code will be labeled with corresponding ``ground-truth'' idiomatic names, since the decompiler often generates variables not present in the original code.
\approach therefore allows an identifier's decompiler-assigned name to be preserved by predicting a special \lt{</identity>} token.

The probability of generating a name is therefore factorized as the product of probabilities of each local decision while generating a sub-token $y_t$:
\begin{equation*}
  p(Y|X) = \prod_{t = 1}^{T} p(y_t | y_{<t}, X),
  \label{eq:decoder:prob}
\end{equation*}
where $X$ denotes the input code, and $Y$ is the full sequence of sub-tokens for all identifiers, and $y_{<t}$ denotes the sequence of sub-tokens before time step $t$.

We model $p(y_t | y_{<t}, X)$ using an LSTM decoder, following the parameterization in~\cite{Luong2015EffectiveAT}.
Specifically, to predict each sub-token $y_t$, at each time step $t$, the decoder LSTM maintains an internal state $\bm{s}_t$ defined by
\[
  \bm{s}_t = f_{\textrm{LSTM}}([\bm{y}_{t - 1} : \bm{v}_{t} : \bm{c}_t], \bm{s}_{t-1}),
\]
where $[:]$ denotes vector concatenation.
The input to the decoder consists of two representations: the embedding vector of the previously predicted name, $\bm{y}_{t - 1}$; and the encoder's representation of the current identifier to be predicted, $\bm{v}_t$.

Our decoder also uses \emph{attention}~\cite{cho2015describing} to compute a context vector $\bm{c}_t$, generated by aggregating contextual information from representations of relevant code elements.
$\bm{c}_t$ is computed by taking the weighted average over encodings of AST nodes and surface code tokens, for each current sub-tokenized name $y_t$.
The decoder's hidden state is then updated using the context vector, incorporating the contextual information into the decoder's state $\tilde{\bm{s}}_t = \bm{W} \cdot [\bm{s}_t : \bm{c}_t]$, where $\bm{W}$ is a weight matrix.
Then, the probability of generating a sub-token ($y_t$) is:
\begin{equation*}
  p(y_t | \cdot) = \frac{\exp \big( \bm{y}^{\intercal}_t \tilde{\bm{s}}_t \big) }{ \sum_{y'}  \exp \big( \bm{y'}^{\intercal} \tilde{\bm{s}}_t \big) }
\end{equation*}

\subsection{Training the Neural Network}
Since \approach is constructed from neural networks, training data is required to learn the weights for each neural component.
Our training corpus is a set $\mathcal{D} = \{ \langle X_i, Y_i \rangle \}$, consisting of pairs of code $X$ and sub-token sequences $Y$, denoting the decoder-predicted sequence of identifier names.
\approach is optimized by maximizing the log-likelihood of predicting the gold sub-token sequence $Y_i$ for each training example $X_i$:
\begin{equation*}
  \sum_{\langle X_i, Y_i \rangle} \log p(Y_i|X_i) = \sum_{\langle X_i, Y_i \rangle} \sum_{t=1}^{|Y_i|} w_t \cdot \log p(y_{i, t}|X_i),
\end{equation*}
where $Y_{i, t}$ denotes the $t$-th sub-token in the decoder's prediction sequence $Y_i$.
As discussed in~\cref{sec:neural_model:decoder}, there are intermediate variables in the decompiled code.
To ensure the decoder network will not be biased towards predicting \lt{</identity>} for other identifiers, we use a tuning weight $w_i$ and set it to 0.1 for sub-tokens that correspond to intermediate variables (and 1.0 otherwise).

\section{Generation of Training Data}
\label{sec:corpus}

Training \approach requires a large corpus of annotated data.
Fortunately, it is possible to create this corpus automatically, starting from a large repository of existing C source code.
At a high level, each entry in our corpus corresponds to a source code function, and consists of the information necessary to train our model.
An entry in the training corpus is illustrated in \cref{fig:corpus}.
Each entry contains three elements: (a) the tokenized code, with variables replaced by an ID that uniquely identifies the variable in the function; (b) the decompiler's AST (\cref{sec:bg:decompilation}) modified to contain the same unique variable IDs; and (c) a lookup table mapping variable IDs to both the decompiler- and developer-assigned names.
It is important to assign a unique variable name to each variable to disambiguate any shadowed variable definitions.
The tokenized code and AST representations are used in both the model's input and output.
The input representation uses the decompiler-assigned names, while the output uses the developer-assigned names.

\label{sec:training-data}
\begin{figure}
  \begin{subfigure}[b]{\columnwidth}
    \tikzstyle{tok} = [rectangle, draw, fill=figgray,
      font=\ttfamily\bfseries\small, minimum height=1.5em]
    \tikzstyle{var} = [tok, draw=figdkblue, fill=figblue]
    \tikzstyle{next} = [draw, -latex']
    \centering
    \begin{tikzpicture}[auto, scale=0.65, every node/.style={transform shape}]
      \node [tok] (1) {for};
      \node [tok, right = 5mm of 1] (2) {(};
      \node [var, right = 5mm of 2] (3) {VAR1};
      \node [tok, right = 5mm of 3] (4) {=};
      \node [right = 5mm of 4] (5) {...};
      \node [var, right = 5mm of 5] (6) {VAR2};
      \node [tok, right = 5mm of 6] (7) {+=};
      \node [var, right = 5mm of 7] (8) {VAR3};
      \node [tok, right = 5mm of 8] (9) {;};
      \node [right = 5mm of 9] (10) {...};
      \path [next] (1) -- (2);
      \path [next] (2) -- (3);
      \path [next] (3) -- (4);
      \path [next] (4) -- (5);
      \path [next] (5) -- (6);
      \path [next] (6) -- (7);
      \path [next] (7) -- (8);
      \path [next] (8) -- (9);
      \path [next] (9) -- (10);
    \end{tikzpicture}
    \caption{Tokenized decompiled code with variable placeholders.}
    \vspace{4mm}
  \end{subfigure}
  \begin{subfigure}[b]{0.5\columnwidth}
    \tikzstyle{var} =[rectangle, draw=figdkblue, fill=figblue, minimum
      width=3em, minimum height=2em, align=left,
      font=\ttfamily\bfseries\small, node distance=.35cm and .1cm]
    \tikzstyle{node} = [var, draw=figdkred, fill=figred, rounded corners]
    \tikzstyle{line} = [draw, -latex']
    \centering
    \begin{tikzpicture}[auto, scale=0.5, every node/.style={transform shape}]
      \node [node] (for) {};
      \coordinate [below = .9cm of for] (belowfor);
      \node [node, left = of belowfor] (11) {};
      \node [node, left = of 11] (16) {};
      \node [node, right = of belowfor] (9) {};
      \node [var, below = of 9] (10) {VAR1};
      \node [node, right = of 9] (6) {};
      \node [var, below = of 6] (7) {VAR2};
      \node [var, right = of 7] (8) {VAR4};
      \node [node, below = of 11] (12) {};
      \node [node, below = of 12] (13) {};
      \coordinate [below = .75cm of 13] (below13);
      \node [var, left = of below13] (14) {VAR2};
      \node [var, right = of below13] (15) {VAR3};
      \node [node, below = of 16] (18) {};
      \node [var, left = of 18] (17) {VAR1};

      \path [line] (for) -- (16);
      \path [line] (for) -- (11);
      \path [line] (for) -- (9);
      \path [line] (for) -- (6);
      \path [line] (6) -- (7);
      \path [line] (6) -- (8);
      \path [line] (9) -- (10);
      \path [line] (11) -- (12);
      \path [line] (12) -- (13);
      \path [line] (13) -- (14);
      \path [line] (13) -- (15);
      \path [line] (16) -- (17);
      \path [line] (16) -- (18);
    \end{tikzpicture}
    \caption{AST with placeholders.}
  \end{subfigure}
  \begin{subfigure}[b]{0.48\columnwidth}
    \centering
    \scriptsize
    \begin{tabular}{lll}
      \toprule
      ID & Decompiler & Developer \\
      \midrule
      1 & \tt{v1} & \tt{ans}\\
      2 & \tt{v2} & \tt{size}\\
      3 & \tt{i}  & \tt{i}\\
      4 & \tt{ptr} & \tt{head}\\
      \bottomrule
    \end{tabular}
    \caption{Variable lookup table.}
  \end{subfigure}
  \caption{Entry in the training corpus.
    Each corresponds to a function and contains (a) tokenized code (b) the AST, both with variables replaced with unique IDs, and (c) a lookup table containing decompiler- and developer-assigned names.\label{fig:corpus}}
\end{figure}

Generating the placeholders and decompiler-chosen names is relatively straightforward.
First, a binary is compiled normally and passed to the decompiler.
Next, for each function, we traverse its AST and replace each variable reference with a unique placeholder token.
Finally, we instruct the decompiler to generate decompiled C code from the modified AST, tokenizing the output.
Thus, we have tokenized code, an AST, and a table mapping variable IDs to decompiler-chosen names.

The remaining step, mapping developer-chosen names to variable IDs, is the core challenge in automatic corpus generation.
Following our previous approach~\cite{Jaffe2018}, we leverage the decompiler's ability to incorporate developer-chosen identifier names into decompiled code when DWARF debugging symbols~\cite{DWARF} are present in the binary.
However, this alone is not sufficient to identify which developer-chosen name maps to a particular variable ID generated in the first step.

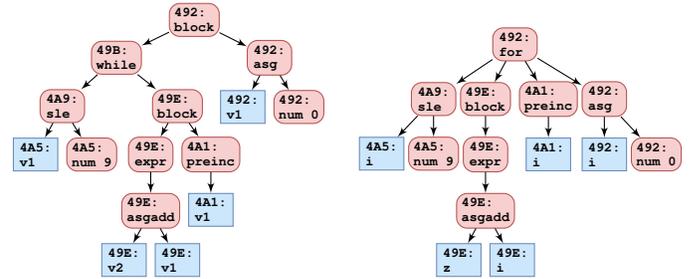
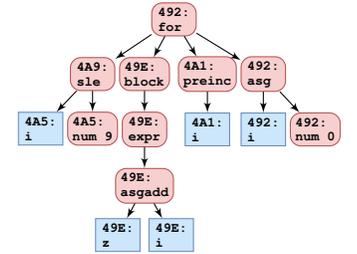
\begin{figure}
  \tikzstyle{var} =[rectangle, draw=figdkblue, fill=figblue, minimum
    width=3em, align=left, font=\ttfamily\bfseries\small, node distance=.5cm
    and .1cm]
  \tikzstyle{node} = [var, draw=figdkred, fill=figred, rounded corners]
  \tikzstyle{line} = [draw, -latex']
    \begin{subfigure}[b]{0.45\columnwidth}
    \centering
    \begin{tikzpicture}[auto, scale=0.56, every node/.style={transform shape}]
      \node [node] (topblock) {492:\\block};
      \coordinate [below = .5cm of topblock] (belowtopblock);
      \node [node, left = 1.25cm of belowtopblock] (while) {49B:\\while};
      \coordinate [below = .75cm of while] (belowwhile);
      \node [node, right = .85cm of belowwhile] (whileblock) {49E:\\block};
      \coordinate [below = .75cm of whileblock] (belowwhileblock);
      \node [node, left = .75cm of belowwhile] (test) {4A9:\\sle};
      \coordinate [below = .75cm of test] (belowtest);
      \node [node, right = of belowtest] (18) {4A5:\\num 9};
      \node [var, left = of belowtest] (17) {4A5:\\v1};
      \node [node, right = of belowwhileblock] (9) {4A1:\\preinc};
      \node [var, below = of 9] (10) {4A1:\\v1};
      \node [node, right = 1.25cm of belowtopblock] (init) {492:\\asg};
      \coordinate [below = .75cm of init] (belowinit);
      \node [var, left = of belowinit] (7) {492:\\v1};
      \node [node, right = of belowinit] (8) {492:\\num 0};
      \node [node, left = of belowwhileblock] (12) {49E:\\expr};
      \node [node, below = of 12] (13) {49E:\\asgadd};
      \coordinate [below = .75cm of 13] (below13);
      \node [var, left = of below13] (14) {49E:\\v2};
      \node [var, right = of below13] (15) {49E:\\v1};

      \path [line] (topblock) -- (while);
      \path [line] (topblock) -- (init);
      \path [line] (while) -- (test);
      \path [line] (while) -- (whileblock);
      \path [line] (whileblock) -- (9);
      \path [line] (init) -- (7);
      \path [line] (init) -- (8);
      \path [line] (9) -- (10);
      \path [line] (whileblock) -- (12);
      \path [line] (12) -- (13);
      \path [line] (13) -- (14);
      \path [line] (13) -- (15);
      \path [line] (test) -- (17);
      \path [line] (test) -- (18);
    \end{tikzpicture}
    \caption{AST without DWARF.\label{fig:tree-while-ast}}
  \end{subfigure}
  ~~~
  \begin{subfigure}[b]{0.45\columnwidth}
    \centering
    \begin{tikzpicture}[auto, scale=0.56, every node/.style={transform shape}]
      \node [node] (for) {492:\\for};
      \coordinate [below = .9cm of for] (belowfor);
      \node [node, left = of belowfor] (11) {49E:\\block};
      \node [node, left = of 11] (16) {4A9:\\sle};
      \node [node, right = of belowfor] (9) {4A1:\\preinc};
      \node [var, below = of 9] (10) {4A1:\\i};
      \node [node, right = of 9] (6) {492:\\asg};
      \node [var, below = of 6] (7) {492:\\i};
      \node [node, right = of 7] (8) {492:\\num 0};
      \node [node, below = of 11] (12) {49E:\\expr};
      \node [node, below = of 12] (13) {49E:\\asgadd};
      \coordinate [below = .75cm of 13] (below13);
      \node [var, left = of below13] (14) {49E:\\z};
      \node [var, right = of below13] (15) {49E:\\i};
      \node [node, below = of 16] (18) {4A5:\\num 9};
      \node [var, left = of 18] (17) {4A5:\\i};

      \path [line] (for) -- (16);
      \path [line] (for) -- (11);
      \path [line] (for) -- (9);
      \path [line] (for) -- (6);
      \path [line] (6) -- (7);
      \path [line] (6) -- (8);
      \path [line] (9) -- (10);
      \path [line] (11) -- (12);
      \path [line] (12) -- (13);
      \path [line] (13) -- (14);
      \path [line] (13) -- (15);
      \path [line] (16) -- (17);
      \path [line] (16) -- (18);
    \end{tikzpicture}
    \caption{AST with DWARF.\label{fig:tree-for-ast}}
  \end{subfigure}
  \caption{Decompiler ASTs for the code in \cref{fig:loops}.
    Hexadecimal numbers indicate the location of the disassembled instruction used to generate the node.
    While the ASTs are different, operations on variables and their offsets are the same, enabling mapping between variables (\ie \lt{v1}$\mapsto$\lt{i} and \lt{v2}$\mapsto$\lt{z}).\label{fig:tree}}
\end{figure}

Specifically, challenges arise because decompilers use debugging information to enrich the decompiler output in a variety of ways, such as improving type information.
Recall from \cref{sec:background} that decompilers often make choices between semantically-identical structures: the addition of debugging information can change which structure is used.
Unfortunately, this means that the difference between code generated with and without debugging symbols is not always an $\alpha$-renaming.
In practice, the format and structure of the code can greatly differ between the two cases.
An example is illustrated in \cref{fig:tree}.
In this example, the first pass of the decompiler is run without debugging information, and the decompiler generates an AST for a \lt{while} loop with two automatically-generated variables named \lt{v1} and \lt{v2}.
Next, the decompiler is passed DWARF debugging symbols and run a second time, generating the AST on the right.
While the decompiler is able to use the developer-selected variable names \lt{i} and \lt{z}, it generates a very different AST corresponding to a \lt{for} loop.

An additional challenge is that there is not always a complete mapping between the variables in code generated with and without debugging information.
Decompilers often generate more variables than were used in the original code.
For example, \lt{return (x + 5);} is commonly decompiled to \lt{int v1; v1 = x + 5; return v1;}.
The decompiled code introduces a temporary variable \lt{v1} that does not correspond to any variable in the original source code.
In this case, there is no developer-assigned name for \lt{v1}, since it does not exist in the original code.
The use of debugging information can change how many of these additional variables are generated.

One solution to these problems proposed by prior work is to post-process the decompiler output using heuristics to \emph{align} decompiler-assigned and developer-assigned names~\cite{Jaffe2018}.
However, this technique can only correctly align 72.8\% of variable names, therefore limiting the overall accuracy of any subsequent model trained on this data.  Instead, we developed a technique that directly integrates with the decompiler to generate an accurate alignment \emph{without using heuristics}.
Our key insight is that while the AST and code generated by the decompiler may change when debugging information is used, \emph{instruction offsets and operations on variables do not change}.
As a result, each variable can be uniquely identified by the set of instruction offsets that access that variable.

For example, in \cref{fig:tree}, although there is not an obvious mapping between the nodes in the trees, the addresses of the variable nodes in the trees have not changed.
This enables us to uniquely identify each variable by creating a signature consisting of the set of all offsets where it occurs.
The variables \lt{v1} and \lt{i} have the signature {\small \tt{\{492,49E,4A1,4A5\}}}, while \lt{v2} and \lt{z} have the signature {\small \tt{\{49E\}}}.
Note that some uses of variables overlap, \eg \lt{v1} (\lt{i}) is summed with \lt{v2} (\lt{z}) in the instruction at offset {\small \tt{49E}}.
This necessitates collecting the full set of variable uses to disambiguate these instances.\footnote{While it is possible for two variable signatures to be identical, we found these collisions to occur very rarely in practice. In these cases we do not attempt to assign names to variables.}

In summary, to generate our corpus we:
1) Decompile binaries containing debugging information.
2) Collect signatures and corresponding developer-assigned names for each variable in each function.
3) Strip debugging information and decompile the stripped binaries.
4) Identify variables by their signature, and rename them in the AST, encoding both the decompiler- and developer-assigned names.
5) Generate decompiled code from the updated AST.
6) Post-process the updated AST and generated code to create a corpus entry.
The final output is a per-binary file containing each function's AST and decompiled code with corresponding variable renamings.

\section{Evaluation}
\label{sec:eval}

We ask the following research questions:
\begin{itemize}
\item RQ1: How effective is \approach at assigning names to variables in
  decompiled code?
\item RQ2: How does each component of \approach contribute to its efficacy?
\item RQ3: How does provenance and quantity of data influence the efficacy of \approach?
\item RQ4: Is \approach more effective than prior approaches?
\end{itemize}

\paragraph{Data Preprocessing}
To answer our first two research questions, we trained \approach on \numfunctions decompiled functions extracted from \numbinaries binaries mined from \textsc{GitHub}.
First, we automatically scraped \textsc{GitHub} for projects written in C.
Next, we modified project build scripts to include debug information when compiling the project, and collected all successfully generated 64-bit x86 binary files.
We then hashed each binary to remove any duplicates. We then passed these binaries through our automated corpus generation system.

Finally, we filtered out any functions that did not have any renamed variables and, for practical reasons, any functions with more than 300 AST nodes.
After filtering, 1,259,935 functions with an average AST size of 77 nodes remained.
These functions were randomly split per-binary into training, development and testing sets with a ratio of 80:10:10.
Splitting the sets per-binary ensures that binary-specific identifiers are not included in both the training and test sets.

\begin{table}
  \caption{Evaluation of \approach.
    Values are percentages, higher accuracy and lower character error rate (CER) are better.\label{fig:accuracy}} \centering
  \begin{tabular}{lrrrrrr}\toprule
    & \multicolumn{2}{c}{\approach} & \multicolumn{2}{c}{Lexical Enc.} &
    \multicolumn{2}{c}{Structural Enc.}\\
    & \multicolumn{1}{c}{Acc.} & \multicolumn{1}{c}{CER}
    & \multicolumn{1}{c}{Acc.} & \multicolumn{1}{c}{CER}
    & \multicolumn{1}{c}{Acc.} & \multicolumn{1}{c}{CER}\\
    \midrule
    Overall           & \bf{74.3} & \bf{28.3} & 72.9 & 28.5 & 64.6 & 37.5\\
    Body in Train     & \bf{85.5} & \bf{16.1} & 84.3 & 16.3 & 75.7 & 25.5\\
    Body not in Train & \bf{35.3} & \bf{67.2} & 33.5 & 67.7 & 26.3 & 76.1\\
    \bottomrule
  \end{tabular}
\end{table}

\paragraph{Evaluation Methodology}
After training, we ran \approach to generate name suggestions on the test data.
We evaluate the accuracy of these predictions, comparing the predicted variable names to names used in the original code (\ie names contained in the debugging information) counting a successful prediction as one exactly matching the original name.
However, there can be multiple, equally acceptable names (\eg \lt{file\_name}, \lt{fname}, \lt{filename}) for a given identifier.
An accuracy metric based on exact match cannot detect these cases.
We therefore use character error rate (CER), a metric that calculates the edit distance between the original and predicted names, then normalizes by the length of the original name \cite{wang2016character}, assigning partial credit to near misses.

Recall from \cref{sec:training-data} that there are often many more variables in the decompiled code than in the original source; these variables will not have a corresponding original name.
In our corpora, the median number of variables in each function is 5, with 3 having a corresponding original name.

Although \approach generates predictions for \emph{all} variables, we do not evaluate predictions on variables that do not have a developer assigned name.
We do this because it is not necessarily incorrect for a renaming system to assign names to variables not present in the original source code.
Recall the example where \lt{return (x + 5);} is decompiled to \lt{int v1; v1 = x + 5; return v1;}.
The name \lt{sum} is likely more informative than \lt{v1}, and it would be unhelpful to penalize a system that suggests this renaming.
However, although renaming in these cases could be helpful, we do not want to overapproximate the effectiveness of our system by claiming any renaming of these variables as correct: it is also possible to assign variables a misleading name that \emph{decreases} the readability of code by obfuscating the purpose of a variable.
For example, suggesting the name \lt{filename} to replace \lt{v1} in the above code would likely be misleading.

\paragraph{Neural Network Configuration}
For our experiments we replicate the neural network configuration of Allamanis \etal~\cite{Allamanis2018}.
We set the size of word embedding layers to be 128.
The dimensionality of the hidden states for the recurrent neural networks used in the encoders is 128, while the hidden size for the decoder LSTM is 256.
For both the sequential and structural encoders, we use two layers of recurrent computation, adding another identical recurrent network to process the decompiled code using the output hidden states of the first layer.
For both \approach and the baseline neural systems, we train each model for 60 epochs.
At testing time, we use beam search to predict the sequence of sub-tokenized names for each identifier (\cref{sec:neural_model:decoder}), with a beam size of 5.

\subsection{RQ1: Overall Effectiveness}
Experimental results are summarized in \cref{fig:accuracy}.
The ``Overall'' row shows the performance of our technique on the full test set and the leftmost column shows the accuracy of \approach.
From this, we can see that \approach can recover \successpct of the original variable names in decompiled code, demonstrating that it is effective in assigning contextually meaningful names to identifiers in decompiled code.

\begin{figure}
  \begin{minipage}{.63\columnwidth}
    \begin{lstlisting}[style=cstyle, morendkeywords={V1,V2,V3}]
void *file_mmap(int V1, int V2)
{
  void *V3;
  V3 = mmap(0, V2, 1, 2, V1, 0);
  if (V3 == (void *) -1) {
    perror("mmap");
    exit(1);
  }
  return V3;
}
    \end{lstlisting}
  \end{minipage}%
  \begin{minipage}{.24\columnwidth}
  \begin{tabular}{lll}
    \toprule
    ID & \approach & Dev.\\
    \midrule
    1 & \lt{fd} & \lt{fd}\\
    2 & \lt{size} & \lt{size}\\
    3 & \lt{buf} & \lt{ret}\\
    \bottomrule
  \end{tabular}
  \end{minipage}
  \caption{Decompiled function (simplified for presentation), \approach variable names, and developer-assigned names.\label{fig:dire-predictions}}
\end{figure}

\Cref{fig:dire-predictions} shows an example renaming generated by \approach.
Here, \approach generates the variable names shown in the ``\approach'' column of the table.
The developer-chosen names are shown in the ``Dev.'' column.
Two of three names suggested by \approach exactly match those chosen by the developer.
Though \approach suggests \lt{buf} instead of \lt{ret} as the replacement for \textcolor{red}{\lt{V3}}, the name is not entirely misleading: \lt{mmap} returns a pointer to a mapped area of memory that can be written to or read from.

Work has shown that large code corpora may contain near-duplicate code across training and testing sets, which can cause evaluation metrics to be artificially inflated~\cite{miltos19duplicate}.
Though our corpus contains no duplicate binaries, splitting test and training sets per-binary still results in functions appearing in both.
A common cause of duplicate functions in different binaries is the use of libraries.
We argue that it is reasonable to allow such duplication since reverse-engineering binaries that link against known (\eg open source) libraries is a realistic use case.

Nevertheless, to better understand the performance of our system, we partition the test examples into two sub-categories: \emph{\textbf{Body in Train}} and \emph{\textbf{Body not in Train}}.
The \emph{Body in Train} partition includes all functions whose entire body matches at least one function in the training set; similarly, the \emph{Body not in Train} set includes only functions whose body does not appear in the training set.
The last two rows in \cref{fig:accuracy} show the performance on these partitions.
\approach performs well on the \emph{Body in Train} test partition (85.5\%).
This indicates that \approach is particularly accurate at name prediction when code has appeared in its training set (\eg libraries, or code copied from another project).
\approach is still able to exactly match 35.3\% of variable names in the \emph{Body not in Train} set, indicating that it still generalizes to unseen functions.

\begin{table}[tb]
\scriptsize
  \caption{Example identifiers from the \emph{Body not in Train} testing partition and \approach's top-$5$ most frequent predictions. }
  \label{tab:exp:example_identifier_and_prediction}
  \centering
  \begin{tabular}{llll}
    \toprule
  \lt{len} & \lt{value} & \lt{new_node} & \lt{bytes_read} \\
    \midrule
    \lt{len} (60\%)   & \lt{value} (28\%) & \lt{node} (48\%)   & \lt{size} (38\%) \\
    \lt{n} (6\%)      & \lt{data} (7\%)   & \lt{child} (31\%)  & \lt{bytes_read} (13\%) \\
    \lt{size} (5\%)   & \lt{val} (3\%)    & \lt{treea} (0.3\%) & \lt{len} (13\%) \\
    \lt{length} (1\%) & \lt{name} (3\%)   & \lt{tree} (0.3\%)  & \lt{cmd_code} (13\%) \\
    \lt{l} (1\%)      & \lt{key} (2\%)    & \lt{root} (0.3\%) & \lt{read} (13\%) \\
    \bottomrule
  \end{tabular}
\end{table}
\Cref{tab:exp:example_identifier_and_prediction} contains example identifiers from the \emph{Body not in Train} test set, along with \approach's most frequent predictions.
We observe that inexact suggested names are often semantically similar to the original names.
\approach also performs best on simple identifiers such as \lt{len} and \lt{value}.
This is because it is difficult to predict the exact name for complex identifiers with compositional names.
However, \approach is still often able to suggest semantically relevant identifiers (\eg \lt{node}, \lt{child}).

\begin{tcolorbox}
  RQ1 Answer: We find that \approach is able to suggest variable names identical to those chosen by the original developer \successpct of the time.
\end{tcolorbox}

\subsection{RQ2: Component Contributions}
\Cref{fig:accuracy} also shows the results for models using only our lexical or structural encoders.
We find that the lexical encoder is able to correctly predict 72.9\% of the original variable names, while a model using the structural encoder is able to correctly predict 64.6\% of the original variable names.
These simpler models still perform well, but by combining them in \approach we are able to achieve even better performance.

\begin{figure}
  \begin{minipage}{\textwidth}
    \begin{lstlisting}[style=cstyle, morendkeywords={V1,V2,V3}]
file *f_open(char **V1, char *V2, int V3) {
  int fd;
  if (!V3)
    return fopen(*V1, V2);
  if (*V2 != 119)
    assert_fail("fopen");
  fd = open(*V1, 577, 384);
  if (fd >= 0)
    return fdopen(fd, V2);
  else
    return 0;
}
    \end{lstlisting}
  \end{minipage}
  \centering
  \begin{tabular}{lllll}
    \toprule
    ID & Lexical & Structural & \approach & Developer\\
    \midrule
    1 & \lt{file} & \lt{fname} & \lt{filename} & \lt{filename}\\
    2 & \lt{name} & \lt{oname} & \lt{mode} & \lt{mode}\\
    3 & \lt{mode} & \lt{flags} & \lt{create} & \lt{is_private}\\
    \bottomrule
  \end{tabular}
  \caption{Decompiled function (simplified for presentation), suggested names, and developer-assigned names.
    The lexical and structural models are unable to correctly predict the name \lt{filename} for variable 1, but \approach can by combining them.\label{fig:predictions}}
\end{figure}

\Cref{fig:predictions} illustrates how \approach can effectively combine these models to improve suggestions.
Here, the placeholders \textcolor{red}{\lt{V1}}, \textcolor{red}{\lt{V2}}, and \textcolor{red}{\lt{V3}} are variables which should be assigned names.
The ``Lexical'', ``Structural'', and ``\approach'' columns show the predictions from each model, and the ``Developer'' column shows the name originally assigned by the developer.
In this example, the lexical and the structural models are unable to predict any of the original variable names, while \approach is able to correctly predict two of the three names.

This example also shows the contributions from each of the submodels.
For example, for \textcolor{red}{\lt{V1}}, the lexical model predicts \lt{file} while the structural model predicts \lt{fname}.
Combining the predicted subtokens generates \lt{filename}, the same name chosen by the developer.
For \textcolor{red}{\lt{V2}}, the lexical and structural models both fail to predict \lt{mode}, but note that the lexical model \emph{does} predict \lt{mode} for \textcolor{red}{\lt{V3}}.
By combining the models, \approach instead correctly predicts \lt{mode} for \textcolor{red}{\lt{V2}}.

\begin{tcolorbox}
  RQ2 Answer: Each component of \approach contributes uniquely to its overall accuracy.
\end{tcolorbox}

\subsection{RQ3: Effect of Data}
\begin{figure*}[t!]
\centering
  \begin{subfigure}[t]{.32\textwidth}
    \includegraphics[width=\textwidth]{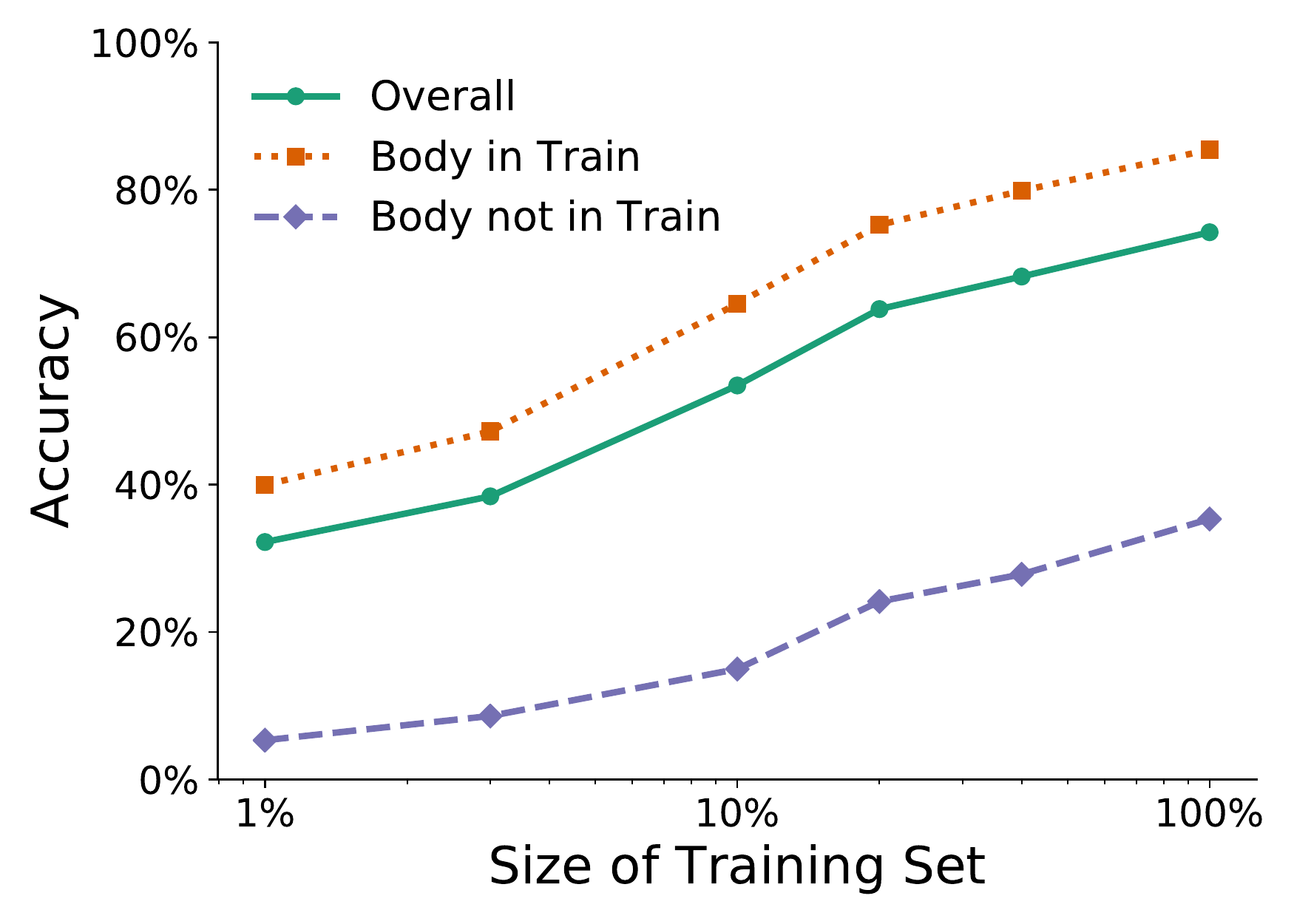}
    \caption{Accuracy of \approach (higher is better).}
    \label{fig:exp:hybrid_model_acc_data_size}
  \end{subfigure}
  \begin{subfigure}[t]{.32\textwidth}
    \includegraphics[width=\textwidth]{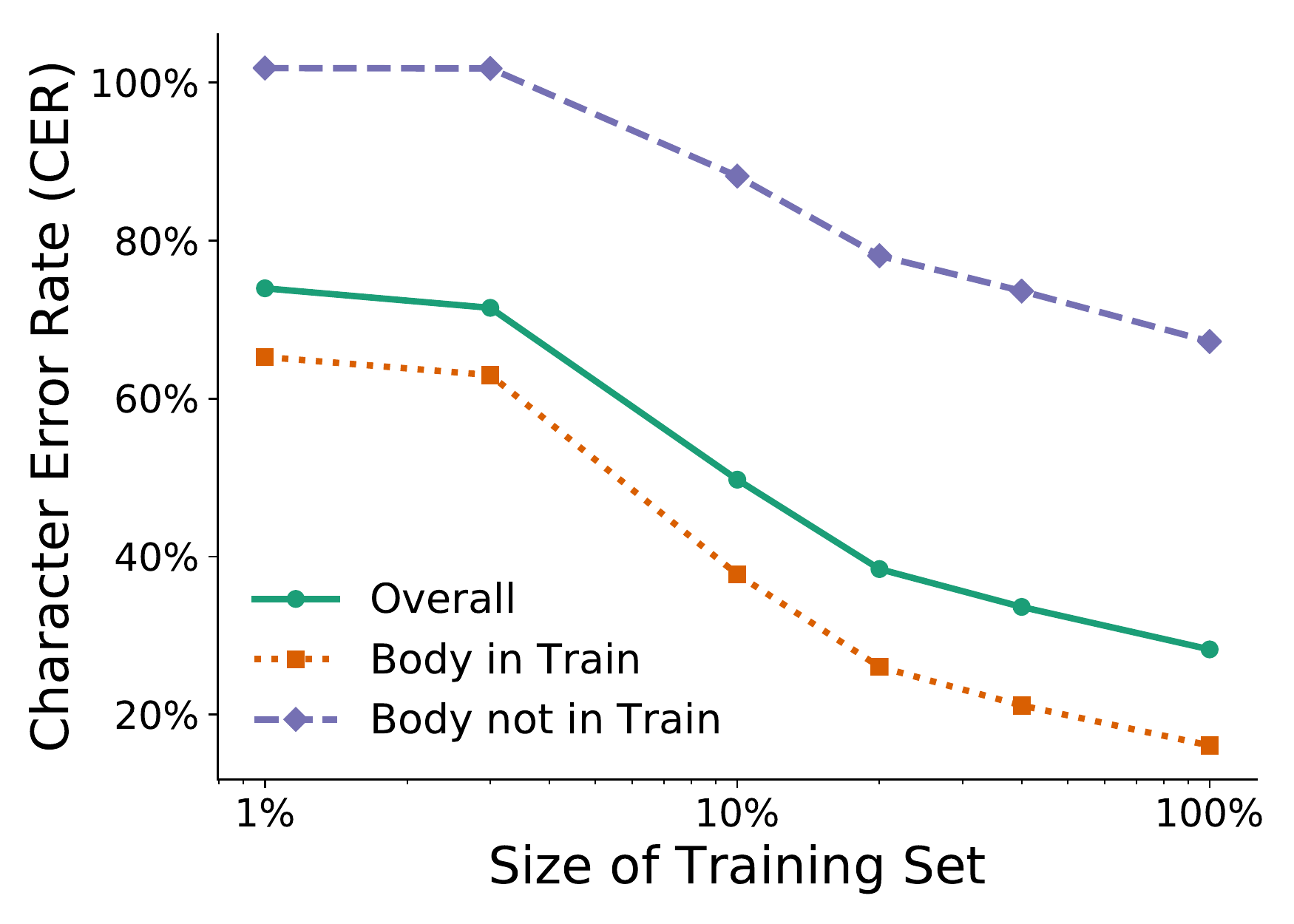}
    \caption{CER of \approach (lower is better).}
    \label{fig:exp:hybrid_model_cer_data_size}
  \end{subfigure}
   \begin{subfigure}[t]{.32\textwidth}
    \includegraphics[width=\textwidth]{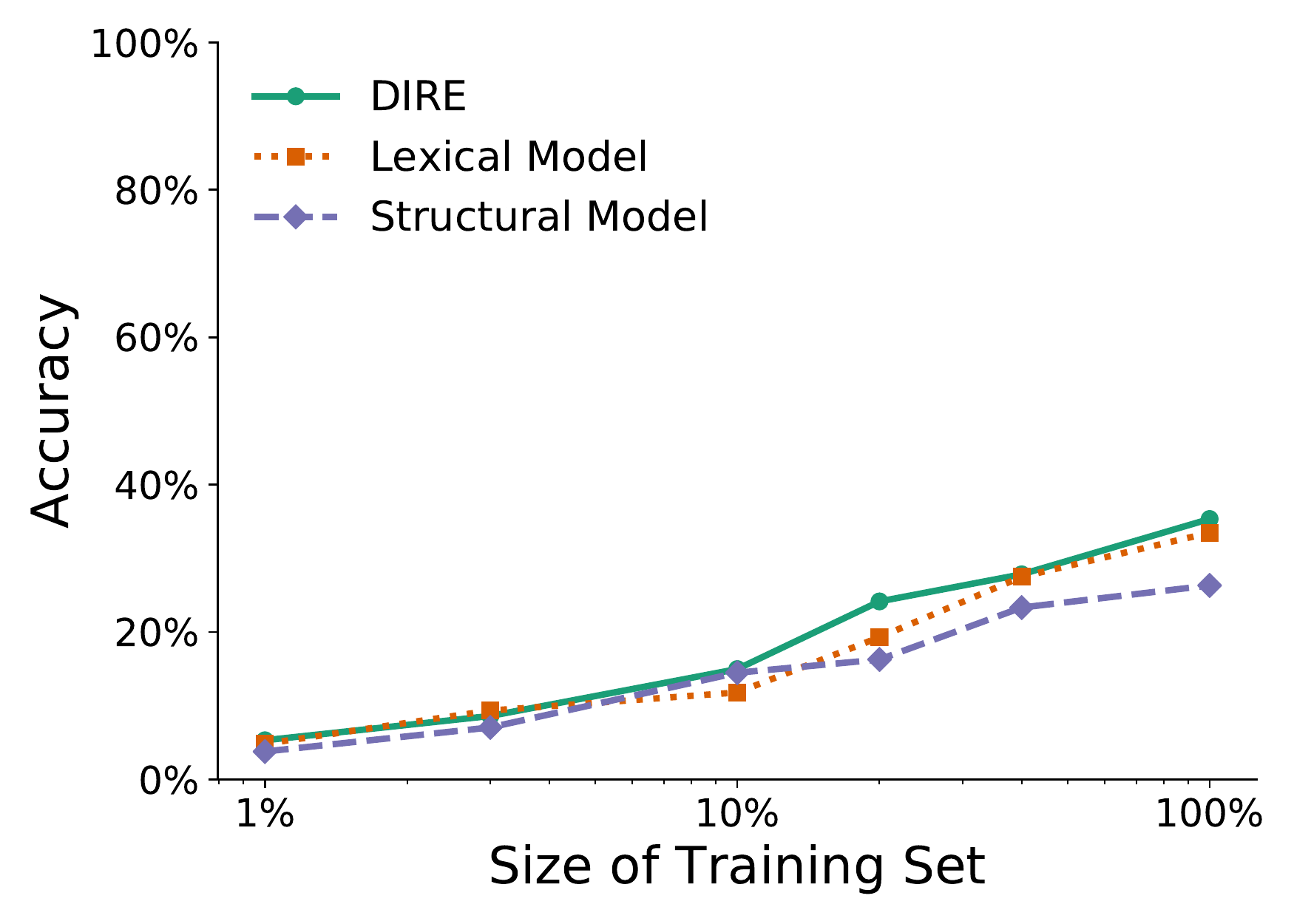}
    \caption{Accuracy of each neural model on the \textit{Body not in Train}
      partition.}
    \label{fig:exp:acc_different_model_data_size}
  \end{subfigure}
   \caption{The impact of training corpus size on the performance of \approach.
     Figures (a) and (b) show how increasing the amount of training data improves the performance of \approach; (c) shows the performance of each of the submodel as training size changes.\label{fig:training-data}}
\end{figure*}

To answer RQ3, we varied the size of the training data and measured the change in performance of our models.
Training data was subsampled at rates of 1\%, 3\%, 10\%, 20\%, and 40\%.
The results of these experiments are shown in \cref{fig:training-data}.

\Cref{fig:exp:hybrid_model_acc_data_size,fig:exp:hybrid_model_cer_data_size} show the change in accuracy and CER of \approach respectively.
The size of the training data is plotted on the $x$-axes, while accuracy and CER are plotted on the $y$-axes.
While \approach has low accuracy on the \emph{Body not in Train} set at the lowest sampling rates, at a 1\% sampling rate it is still able to correctly select names over 40\% of the time for the \emph{Body in Train} test set, suggesting that it is possible to use much less data to train a model if the target application is reverse engineering of libraries rather than binaries in general.

Note however that the CER of \approach is still high at low sampling rates.
This implies that in the cases where \approach selects an incorrect variable name the chosen name is quite different from the correct name.
Sampling at a higher rate dramatically decreases the CER, allowing for namings that are closer the developers' choices.
At a sampling rate of 40\%, \approach comes quite close to the performance of the model trained on the full training set, with an overall accuracy of 68.2\% (vs. 74.2\%) and a CER of 33.6\% (vs. 28.2\%).

\Cref{fig:exp:acc_different_model_data_size} shows the effect of training set size on the performance of \approach and its component neural models on the \emph{Body not in Train} test set.
Note how at sampling rates at or below 10\% the models have similar performance.
In cases where there is little training data, training time can be further reduced by using only one of the two submodels.

\begin{tcolorbox}
  RQ3 Answer: \approach is data-efficient, performing competitively using only 40\% of the training data.
  \approach is also robust, outperforming the lexical and structural models in most sub-sampling cases.
\end{tcolorbox}

\subsection{RQ4: Comparison to Prior Work}
\begin{figure}
  \begin{subfigure}[t]{.5\columnwidth}
    \begin{lstlisting}[style=cstyle]
long gray(unsigned a1,
          int a2) {
  unsigned v3, v4;
  int v5;
  if (a2 >= 0)
    return a1 ^ (a1 >> 1);
  v5 = 1;
  v4 = a1;
  while (1) {
    v3 = v4 >> v5;
    v4 ^= v4 >> v5;
    if (v3 <= 1 ||
        v5 == 16)
      break;
    v5 *= 2;
  }
  return v4;
}
    \end{lstlisting}
    \caption{Hex-Rays.}
    \label{fig:hexrays-no-debug}
  \end{subfigure}
  \begin{subfigure}[t]{.49\columnwidth}
    \begin{lstlisting}[style=cstyle]
void gray() {
  unsigned v0;
  int v1;
  unsigned i, v3;
  int x;
  if (v1 < 0) {
    x = 1;
    v3 = v0;
    while (1) {
      i = v3 >> x;
      v3 ^= v3 >> x;
      if (i <= 1 ||
          x == 16)
        break;
      x *= 2;
    }
  }
}
    \end{lstlisting}
    \caption{Hex-Rays w/ \debin.}
    \label{fig:hexrays-debin}
  \end{subfigure}

  \caption{Effects of incorrect debugging information on decompiler output.
    The \lt{gray} function computes the Gray code of \lt{a1} in \lt{a2} bytes~\cite{Press1992}.
    On the left, (a) is the output of Hex-Rays without debugging symbols; it is able to correctly identify the arguments and return type.
    On the right, (b) is the output with incorrect DWARF information generated by \debin: note missing arguments, \lt{return} statements, and incorrect type.}
  \label{fig:bad-debug}
\end{figure}

To answer RQ4, we compare to our prior work~\cite{Jaffe2018} and to \debin~\cite{DEBIN}, the state-of-the-art technique for predicting debug information directly from binaries.

In our earlier work, which used a purely-lexical model based on statistical machine translation (SMT), we were able to exactly recover 12.7\% of the original variable names chosen by developers.
In contrast, \approach is able to suggest identical variable names 74.3\% of the time.
We attribute this improvement to two factors: 1) the improved accuracy of our corpus generation technique, and 2) the use of a model that incorporates both lexical and structural information.

To better understand the performance of \approach, we also compare to \debin, a different approach to generating more understandable decompiler output.
\debin uses CRFs to learn models of binaries and directy generate DWARF debugging information for a binary, which can be used by a decompiler such as Hex-Rays.

The debugging information generated by \debin contains predicted identifiers, types, and names.
To choose a variable name, \debin proceeds in two stages: it predicts which memory locations correspond to function-local arguments and variables, and then predicts names for the variables it identified.
In contrast, \approach leverages the decompiler to identify function offsets and local variables.

Building on top of the decompiler helps \approach maintain the quality of pseudocode output.
An example is shown in \cref{fig:bad-debug}, which contains a C function for converting between a number \lt{a1} and its Gray code representation in \lt{a2} bits~\cite{Press1992}.
\Cref{fig:hexrays-no-debug} shows the output of Hex-Rays when passed a binary with no debug information.
Although these variables do not have meaningful names, it is clear that \lt{gray} is a function that takes two arguments and returns a \lt{long}.

\Cref{fig:hexrays-debin} shows the output of Hex-Rays using debugging information generated using \textsc{Debin}'s bundled model.\footnote{https://files.sri.inf.ethz.ch/debin\_models.tar.gz, accessed April 10, 2019}
We observe that \textsc{Debin} does not accurately recover variable names in this case, perhaps since its model was trained on a different set of code.

However, this example also surfaces a fundamental limitation of the \textsc{Debin} approach: both the inferred structure and the types of the variables in the program have changed.
This occurs because Hex-Rays prioritizes debugging information over its own analyses and heuristics.
In this case, the debugging information generated by \debin does not indicate a return value of the \lt{gray} function nor any arguments, misleading the decompiler.
By starting at the point shown in \cref{fig:hexrays-no-debug} \approach maintains structure and typing even in the presence of incorrect predictions.

To evaluate our performance compared to \debin, we trained it on binaries in our dataset.
Due to time restrictions, we found it impractical to train \debin on the full dataset.
For a fair comparison, we instead subsampled our training set at 1\% and 3\% and trained both \debin and \approach on these sets.\footnote{The 3\% subsampling we used is a slightly larger training set than the 3,000 binaries used to train \debin in the original paper~\cite{DEBIN}.}
After training, we ran \debin on binaries in our test set, extracted names using our corpus generation pipeline, and measured the accuracy of predictions.
Our results are shown in \cref{tab:debin-comparison}.

\begin{table}
  \caption{Comparison of \approach and \debin trained on 1\% and 3\% of our full corpus of \numbinaries binaries.
    All accuracy values are percentages, higher accuracy is better.
    \label{tab:debin-comparison}} \centering
  \begin{tabular}{lrrrr}\toprule
    & \multicolumn{2}{c}{1\% of corpus} & \multicolumn{2}{c}{3\% of corpus}\\
    & \multicolumn{1}{c}{\approach} & \multicolumn{1}{c}{\debin}
    & \multicolumn{1}{c}{\approach} & \multicolumn{1}{c}{\debin}\\
    \midrule
    Training Time (h) &   1.8   & 13.3 &   6.1   & 17.2\\
    \midrule
    Accuracy -- Overall           & 32.2 & 2.4 & 38.4 & 3.9 \\
    Accuracy -- Body in Train     & 40.0 & 3.0 & 47.2 & 4.8 \\
    Accuracy -- Body not in Train & 5.3 & 0.6 & 8.6 & 0.7 \\
    \bottomrule
  \end{tabular}
\end{table}

We find that \approach is able to outperform \debin at all sampling sizes.
When trained on 1\% of the corpus \approach is able to exactly recover 32.2\% of all identifiers, while \debin recovers 2.4\%.
On the 3\% partition, \approach is able to recover 38.4\% of names, while \debin is able to recover 3.9\%.
The lower performance of \debin we observed could be attributed to compound error: in addition to variable names themselves, \debin must predict what memory locations correspond to variables.
If a memory location is not predicted to be a variable, \debin cannot assign it a name.

We also note that we were able to train \approach much faster than \debin, although \approach is GPU-accelerated, while \debin as distributed is limited to execution on the CPU.

\begin{tcolorbox}
  RQ4 Answer: \approach is a more accurate and more scalable technique for variable name selection than other state-of-the-art approaches.
\end{tcolorbox}

\section{Threats To Validity}

When collecting code and binaries to generate our corpus, we did no filtering of the repositories beyond ensuring that they were written in C and built.
It is possible that the code we collected does not accurately represent the types of binaries that are typically targets of reverse-engineering effort.

Additionally, we did not experiment with binaries compiled with optimization enabled, nor did we experiment with intentionally obfuscated code.
It is possible that \approach does not perform as well on these binaries.
However, reverse engineering of these binaries is a general challenge for decompilers, and we do not believe that our technique applies exclusively to the test code we experimented with.

Although we have found that it is possible to uniquely identify variables in Hex-Rays based on the code offsets where it is accessed, we have found that other decompilers do not have this property.
In particular, our approach did not work well with the newly released Ghidra decompiler~\cite{ghidra}.
One of the primary causes is the way that Hex-Rays and Ghidra utilize debug symbols to name variables.
Hex-Rays uses debug symbols in a very straight-forward manner, and generally does not propagate local names outside of their function.
Ghidra, however, will actually propagate variable names at some function calls.
For example, if an unnamed variable is passed as an argument to a function whose parameter has a name, in some cases Ghidra will rename the variable to match the parameter's name.
This behavior is problematic for corpus generation because it does not reflect the developer's intended names.

A new approach for corpus generation would be required for compatibility with Ghidra, but Ghidra's open-source nature (as opposed to Hex-Rays' closed model) allows potential modification of the decompiler, including disabling the problematic propagation of names at function calls.
We leave Ghidra integration to future work.

\section{Conclusion}
Semantically meaningful variable names are known to increase code understandability, but they generally cannot be recovered by decompilers.
In this paper, we proposed the \expandedname (\approach), a novel, probabilistic technique for variable name recovery which uses both lexical and structural information.
We also presented a technique for generating corpora suitable for training \approach, which we used to generate a corpus from \numbinaries unique x86-64 binaries.
Our experiments show that \approach is able to predict variable names identical to the names used in the original source code up to \successpct of the time.

\section{Acknowledgments}
This material is based upon work supported in part by the Software Engineering Institute (LINE project 6-18-001) and National Science Foundation (awards 1815287 and 1910067).
We also gratefully acknowledge hardware support from the NVIDIA Corporation.
Computation for this research was also supported in part by the Pittsburgh Supercomputing Center and a gift of AWS credits from Amazon.
Thanks to both Prem Devanbu and members of the CERT Division at the Software Engineering Institute for helpful feedback on earlier drafts.

\newpage
\balance
\bibliographystyle{IEEEtran}
\bibliography{references}

\end{document}